\documentclass[aps,prl,twocolumn,a4paper,10pt,notitlepage,footinbib,superscriptaddress,longbibliography]{revtex4-1}
\usepackage[english]{babel}
\usepackage{lmodern}
\usepackage[utf8]{inputenc}
\usepackage{endnotes}
\usepackage{amssymb,amsmath,amsfonts}
\usepackage{textcomp}
\usepackage{braket}
\usepackage{ulem}
\usepackage{soul}
\usepackage{graphicx}
\usepackage{dcolumn}
\usepackage{bm}
\usepackage{xcolor}
\usepackage{soul}
\usepackage{xr}
\usepackage{cleveref}

\definecolor{crimson}{rgb}{0.75686,0,0.262745}
\definecolor{saphire}{rgb}{0.0,0.196,0.372549}
\definecolor{plum}{rgb}{0.50588,0.007843,0.3843137}

\begin{document}


\title{Controlling flow patterns and topology in active emulsions}


\author{Giuseppe Negro$^{*}$}
\affiliation{Dipartimento di Fisica, Università degli Studi di Bari and INFN, Sezione di Bari, via Amendola 173, Bari, I-70126, Italy}
\author{Louise C. Head$^{*}$}
\affiliation{SUPA, School of Physics and Astronomy, University of Edinburgh, Peter Guthrie Tait Road, Edinburgh, EH9 3FD, UK}
\author{Livio N. Carenza} 
\affiliation{Faculty CS Physics, Koc University, Rumelifeneri Yolu 34450
Sariyer, Istanbul,  Turkey}
\author{Tyler N.~Shendruk}
\affiliation{SUPA, School of Physics and Astronomy, University of Edinburgh, Peter Guthrie Tait Road, Edinburgh, EH9 3FD, UK}
\author{Davide Marenduzzo}
\affiliation{SUPA, School of Physics and Astronomy, University of Edinburgh, Peter Guthrie Tait Road, Edinburgh, EH9 3FD, UK}
\author{Giuseppe Gonnella}
\affiliation{Dipartimento di Fisica, Università degli Studi di Bari and INFN, Sezione di Bari, via Amendola 173, Bari, I-70126, Italy}
\author{Adriano Tiribocchi}
\affiliation{Istituto per le Applicazioni del Calcolo, Consiglio Nazionale delle Ricerche, via dei Taurini 19, Roma, 00185, Italy}
\affiliation{INFN "Tor Vergata" Via della Ricerca Scientifica 1, 00133 Roma, Italy\\
$^*$Equal contribution}

\begin{abstract}
Active emulsions and liquid crystalline shells are intriguing and experimentally realisable types of topological  matter. Here we numerically study the morphology and spatiotemporal dynamics of a double emulsion, where one or two passive small droplets are embedded in a larger active droplet. We find activity introduces a variety of rich and nontrivial nonequilibrium states in the system. First, a double emulsion with a single active droplet becomes self-motile, and there is a transition between translational and rotational motion: both of these regimes remain defect-free, hence topologically trivial.
Second, a pair of particles nucleate one or more disclination loops, with conformational dynamics resembling a rotor or  chaotic oscillator, accessed by tuning activity. In the first state a single, topologically charged, disclination loop powers the rotation. In the latter state, this disclination stretches and writhes in 3D, continuously undergoing recombination to yield an example of an active living polymer. 
These emulsions can be self-assembled in the lab, and provide a pathway to form flow and topology patterns in active matter in a controllable  way, as opposed to bulk systems that typically yield active turbulence. 
\end{abstract}


\maketitle

\section{Introduction}


Active nematics are  soft matter and biophysical systems~\cite{marchetti,ramaswamy} that  can be self-assembled experimentally by using mixtures of cytoskeletal polymers and molecular motors, such as actin fibres and myosin, or microtubules and kinesin~\cite{sanchez2012,kumar2018}. In addition,  experimental and theoretical work has shown that even confluent monolayers of deformable cells can, under certain conditions, be described as active nematics~\cite{doostmohammadi2018,armengol2023,chiang2023,Armengol2024}. Whilst most studies of active nematics have been performed in the bulk, theory and simulations suggest that 2 and 3-dimensional active droplets and 3D active nematic shells give rise to rich phenomenology, ranging from spontaneous motility due to internal self-sustained active flow patters, to the self-organisation into cell-like morphologies~\cite{tjhung,tjhung2015,ruske2021,carenza2019,Hoffmann2022,Hoffmann2023,Nejad2023}. These systems are becoming increasingly relevant to materials science as well, due to advances that allow experimental studies of 3D active nematics in the bulk or under confinement~\cite{duclos2020,zhang2016}.

A particularly interesting confining geometry is that of double (or multiple) emulsions, where one or more passive isotropic droplets (or cores) are embedded in a larger active liquid crystalline droplet. Experimentally, these systems can be realised by water-oil-water emulsions produced by microfluidic techniques~\cite{MaaasPRL}; when there is a single passive core, the system is equivalent to a liquid crystalline thick shell~\cite{MaaasPRL,review_autonomous_materials,carenza2022,negro2023}.  Active droplets and double active emulsions provide a remarkable example of tunable 3D topological active matter. Indeed, even a single 3D active droplet with tangential anchoring is topologically non-trivial, as the Poincar\'e-Hopf theorem prescribes that it must  accommodate surface defects with topological charge summing to $+2$, the Euler characteristic of a sphere~\cite{carenza2019}. With normal anchoring, it is the 3D {\it interior} of the droplet which is topologically non-trivial, such that there must be at least one point topological defect or, equivalently, a topologically  charged disclination loop~\cite{binysh2020}. Because charge  is only conserved modulo $2$ in 3D nematics~\cite{mermin1979}, multiple disclination patterns are possible in principle, provided that they are globally non-trivial --- i.e., they are not equivalent to a defect-free state. 

For the case of double emulsion, we focus on normal anchoring --- both on the internal droplets and on the external one. Thus, each passive core is equivalent to a point (or hedgehog) charge, and this must be accounted for when computing the global topological charge. Thus, a double emulsion with a single core, i.e. a thick active nematic shell, is topologically trivial, such that the sum of the charges of any disclination loop must be equal to $0$ modulo $2$. Instead, a multiple emulsion with two passive cores is topologically non-trivial, such that the topological charge of the internal pattern equals $1$,  modulo $2$. More  generally, emulsions with an odd number of passive cores are topologically trivial, while those with an even number are topologically non-trivial. Therefore, the emulsion geometry can be used to control the topology of the emerging patterns in a simple way. Because flow field and orientational order are strongly coupled in active nematics --- as elastic deformations drive flow, which in turn reorients the nematic --- emulsion geometry and strength of activity also offer an appealing pathway to select macroscopic flow patterns. 



Here, we use computer simulations to study the spatiotemporal dynamics and morphology of multiple active/passive emulsions with normal anchoring. Our analysis characterises how the emulsion topology combines with activity to affect the emergent physics. 
Focussing on the cases of extensile activity and a single or a pair of passive cores, we find rich physical behaviour, encompassing controllable nonequilibrium spatiotemporal motility patterns and nontrivial disclination topologies. First, we find that an emulsion with a single passive droplet within a larger active droplet can become self-motile in a defect-free configuration, with an activity-driven transition between a translating and a rotating regime. \textcolor{black}{Second, an emulsion with two passive cores, at low activity, induces a charged disclination loop that behaves as a self-assembled spinner. On the other hand,  at large activity, the internal flow loses coherence and  chaotic dynamic ensues}. Corresponding to this transition, the disclination line in the system significantly stretches and writhes, continually supporting topologically-protected self-recombination. This yields states with odd numbers of charged disclination loops, and integer numbers of neutral loops. 
The observed charged disclination lines are qualitatively different from  conventional  Saturn rings that hug colloidal passive particles embedded in a nematic liquid crystal~\cite{copar2013}. In conventional Saturn rings, the local director profile is always that of a defect with topological charge $-1/2$ along the whole loop; however, in the observed charged loop,  the local profile transforms twice from a $+1/2$ to a $-1/2$ local defect profile, where the $-1/2$ regions tend to localise at passive droplet surfaces. 

\begin{figure}[t!]
\includegraphics[width=1.0\columnwidth]{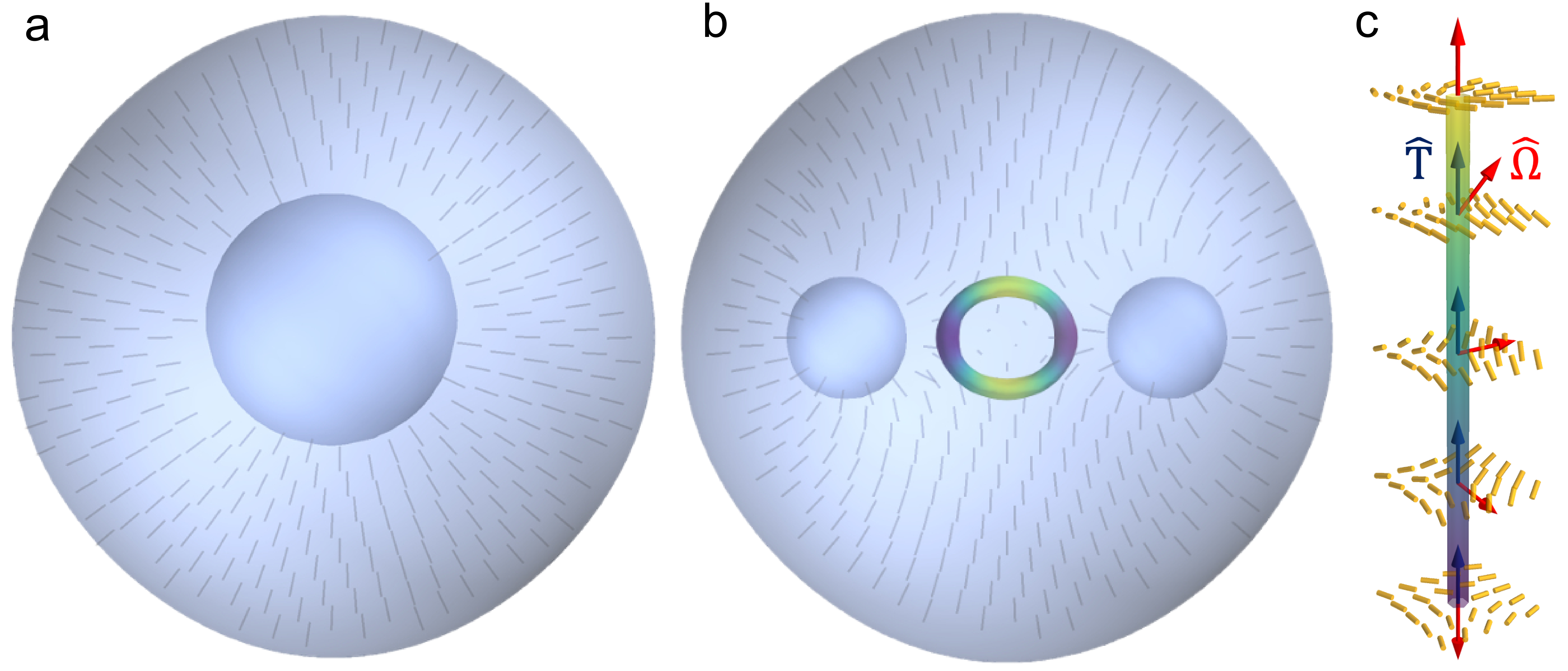}
\caption{\textbf{Equilibrium structures of passive double emulsions with homeotropic anchoring.} (a) Single-core double emulsion. (b) Two-core double emulsion. 
Headless vectors represent the director field. 
The disclination line (see Methods section) has been coloured accordingly to $\cos\beta=\bm{\Omega}\cdot{\bf T}$, where ${\bm{\Omega}}$ is the rotation vector and ${\bf T}$ the disclination line tangent (c).  This allows for the identification of different local director profiles, such as twist-type ($\cos\beta=0$), $+1/2$ comet-shaped ($\cos\beta=1$) and $-1/2$ trefoil-shaped ($\cos\beta=-1$).}
\label{fig1}
\end{figure} 


\begin{figure*}[t!]
\includegraphics[width=2.0\columnwidth]{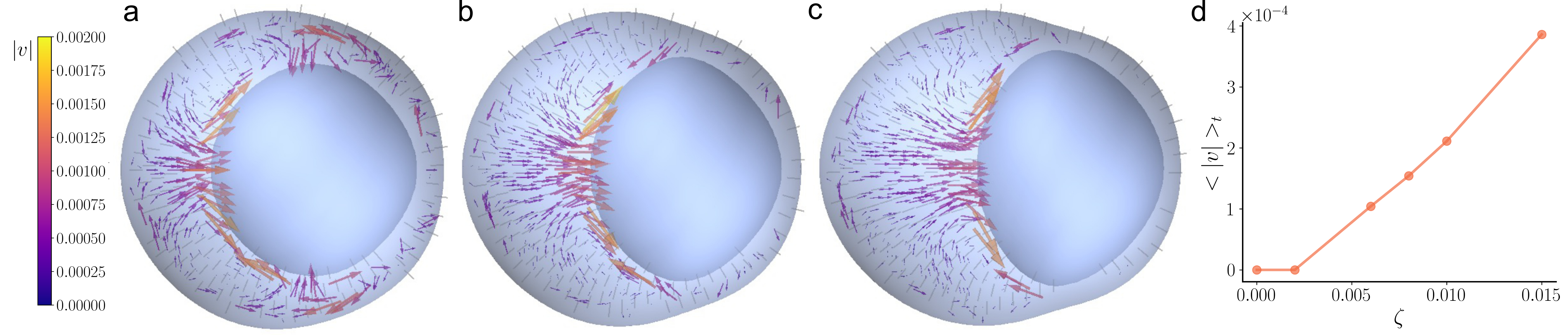}
\caption{\textbf{Extensile double emulsion with homeotropic anchoring: Translational motion.} (a)-(c) For $\zeta=10^{-2}$, the encapsulated drop acquires motion (panels correspond to $t=10^5$, $t=1.5\times 10^{5}$, $t=7\times 10^{5}$ respectively), driven by the spontaneous flow  produced by the active gel. In steady state (c), the emulsion translates along the direction imposed by that of the inner drop. Coloured arrows indicate the velocity field while the headless ones denote the director. (d) Center of mass speed, averaged over time, for different values of activity. For $\zeta\lesssim 10^{-3}$ no spontaneous motion is observed.}
\label{fig2}
\end{figure*} 
\section{Results}
\noindent{\bf Equilibrium states of a passive double emulsion with homeotropic anchoring.} We begin by describing the equilibrium configuration of a passive liquid crystalline double emulsion.  This is initialised by encapsulating a spherical isotropic droplet (core) of radius $R_1$ within an external nematic droplet of radius $R_2>R_1$, both placed at the center of a cubic lattice of linear size $L=128$ (see Fig.\ref{fig1}a). This double emulsion is, in turn, surrounded by further passive fluid acting as a solvent. In these simulations, $R_1=16$ and $R_2=32$ lattice sites. The resulting nematic shell has uniform radial orientation due to the perpendicular anchoring imposed at both fluid interfaces.  A suitable dimensionless number measuring the strength of such anchoring, controlled by the constant $W$, relative to the bulk elasticity, set by $\kappa$, is the ratio $WR_l/(\kappa l_i)$, where $R_l=R_2-R_1$ (see also the section Methods for further details) and $l_i$ is the interfacial thickness of the droplet (equal to a few lattice sites in our simulations). 
 In agreement with the theory of 3D nematic defects~\cite{lavrentovich2001,mermin1979}, the perpendicular anchoring of the liquid crystal at the droplet interface nucleates an imaginary defect of topological charge $1$ (hedgehog defect) located in the center of the inner drop.  \textcolor{black}{Such a design is obtained using a multi-phase model~\cite{NegroScienceAd,nat_tir1,nat_tir2}, in which each droplet is represented by a different phase field. Droplets coalescence is avoided employing a repulsion term between these phase fields in the free-energy (see Methods). }

On  relatively short time scales, the phase fields and the nematic orientation tensor ${\bf Q}$ (see Methods) relax towards the equilibrium values set by minimising the free energy, leading to the double emulsion shown in Fig.\ref{fig1}(a). 

{\bf Translational regime.} For low activity $\zeta$, a  transitional regime occurs (Fig.\ref{fig2} and Supplementary Movie 1).  The entire emulsion acquires a spontaneous motion along a rectilinear trajectory and droplet rotations are absent in this regime. 
The motility mechanism is due to the interplay between anchoring-induced pattern and active flow, and is subtly distinct from the mechanism leading to spontaneous flow in the bulk, which relies on the well-known generic hydrodynamic instability of bend fluctuations of active extensile dipoles~\cite{marchetti,ramaswamy}. In this case, the anchoring leads to a splay-dominated pattern (as in the passive limit). If the internal core is exactly at the centre of the active droplet, this state is quiescent and the double emulsion is nonmotile. However, if the core is displaced, the splay patterns are no longer balanced and, for sufficiently large extensile activity $\zeta$, the active flow renders the emulsion self-motile.  The direction of motion is along the initial displacement (Fig.\ref{fig2}), here along the positive $x$ axis (Fig.\ref{fig2}a), and in general chosen by spontaneous symmetry breaking. In steady state (Fig.\ref{fig2}c),  \textcolor{black}{the flow within the active phase, outside of the passive core,} consists of two large counter-rotating vortices, located in the upper and lower part of the emulsion (Fig.\ref{fig2}a-c). Bend distortions are accompanied by a large slip velocity at the rear of the inner core (Fig.\ref{fig2}c). The internal flow squeezes and displaces the inner core further forward and concurrently stretches the shell along the same direction (Fig.\ref{fig2}b), such that the double emulsion attains a pear-like shape in steady state (Fig.\ref{fig2}c). 

Such translational dynamics are generally observed for  $10^{-3}\lesssim \zeta\lesssim 1.5\times 10^{-2}$, with  speed increasing nearly linearly with activity (Fig.\ref{fig2}(d)).
For lower values of $\zeta$, the active stress is not sufficient to overcome the resistance to deformations of the liquid crystals mediated by the elastic constant $\kappa$. This balance can be quantified in terms of an active Ericksen number, defined as $Er=\zeta (R_2-R_1)^2/\kappa$, which is $Er\simeq 1$ for $\zeta=10^{-3}$. On a general basis, if $Er\leq 1$ the droplet remains at rest, whereas it acquires self-propulsion for higher values. 

\begin{figure*}[t!]
\includegraphics[width=2.0\columnwidth]{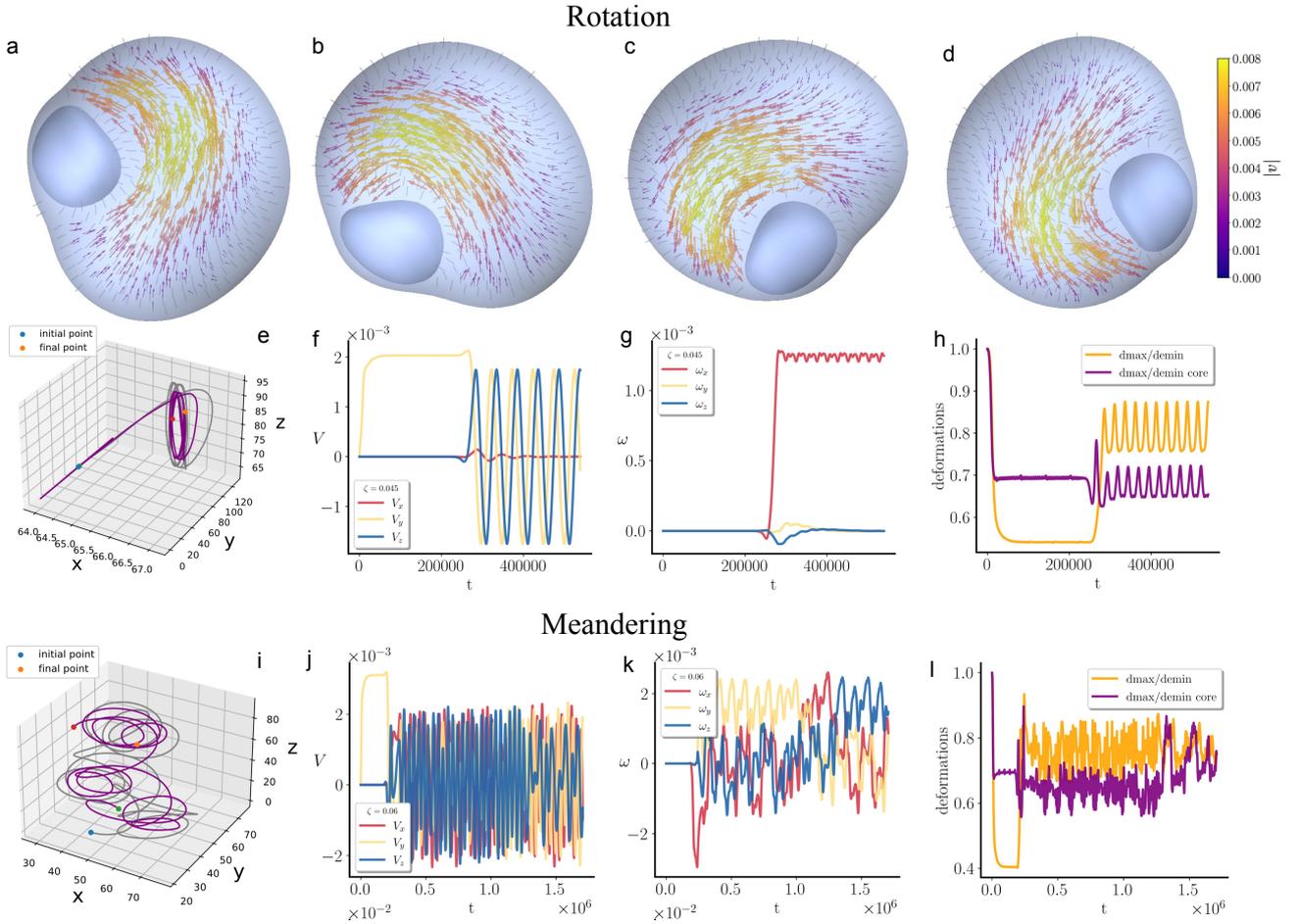}
\caption{\textbf{Extensile double emulsion with homeotropic anchoring: Rotation and Meandering regime.} (a-d) Time sequence of the sum of the phase fields (blue region), fluid velocity  (vectors) and director field (headless vectors) of the emulsion at the steady state in the rotating regime (panels correspond to $t=4\times 10^5$, $t=4.1\times 10^{5}$, $t=4.2\times 10^{5}$ and $t=4.3\times 10^{5}$ respectively), for $\zeta=0.045$. 
(e) Trajectory of the center of mass of the emulsion for the same case. Panels (f-h) show  the time evolution of  speed, angular velocity, and the deformation parameter, respectively. 
(i-l) Trajectory, speed, angular velocity and time evolution of  deformations,  for $\zeta=0.06$. In this case the dynamics is characterised by a chaotic-like behaviour.}
\label{fig3}
\end{figure*} 
With respect to 
previous studies of self-propelled droplets \cite{tjhung,hawkins,Nejad2023,ruske2021}, our design benefits from the presence of the encapsulated core that explicitly dictates the direction of motion, thus considerably facilitating the dynamic control of the emulsion.
Increasing $\zeta$ dramatically changes the picture illustrated so far. The next sections describe two further motility regimes emerging for higher values of activity. 

{\bf Rotating regime}. \textcolor{black}{Increasing activity  ($\zeta\geq 4.5 \times 10^{-2}$, or $Er\gtrsim 44$)  a new self-motility regime is encountered (Fig.\ref{fig3} and Supplementary Movie 2), where the encapsulated droplet is found to display a persistent periodic motion along a circular path (Fig.\ref{fig3}a-d)}. The active emulsion initially exhibits a transient behaviour akin to that observed in the translational phase, where the motion is associated with a vortex pair. 
However, the high value of $\zeta$ destabilises this pattern, leading to the merging of the vortices into a single vortex whose direction of rotation is chosen at random, depending on the spontaneous symmetry breaking of the director orientation. Indeed, unlike the previous translational regime (Fig.\ref{fig2}), here the activity  bends the director profile almost everywhere in the shell (except at the interfaces where \textcolor{black}{strong} homeotropic anchoring is enforced), pointing perpendicularly to the local fluid velocity.
The active flow concurrently fosters  the circular motion of the internal droplet and the periodic dynamics of the double emulsion, an effect that can be quantitatively captured  by tracking position, speed and angular velocity of its center of mass (Fig.\ref{fig3}e,f,g). At late times, the core center of mass exhibits long-lasting oscillations of constant amplitude and frequency. 
This  destabilises core and shell shape, which periodically elongate and contract albeit not simultaneously (Fig.\ref{fig3}h). The degree of morphological deformation is quantified by the ratio of the lengths of the major axis $d_{max}$ and minor axis $d_{min}$. Interestingly, the \textcolor{black}{core} is subject to  larger shape deformations than the shell, an effect due to  the sheared structure of the fluid vortex. Despite being a fully 3D process, the rotation remains essentially confined to a plane (the $xy$ plane in Fig.\ref{fig3}e). 

{\bf Meandering regime}. Further increasing activity \textcolor{black}{($\zeta\geq 0.55\times 10^{-2}$, or $Er\gtrsim 54 $)} leads to chaotic-like dynamics (see Supplementary Movie 3 and Fig.\ref{fig3}i-l), closely resembling the meandering motion experimentally observed in self-propelled water-oil-water emulsions hosting a nematic liquid crystal within the shell \cite{MaaasPRL}. In this meandering regime, the higher value of activity destabilises the in-plane structure
observed for the rotating regime (Fig.\ref{fig3}e-h). This  eventually yields a fully three dimensional state in which the double emulsion trajectory grows noisy and the center of mass speed displays rapid oscillations
(Fig.\ref{fig3}j,k). While the oscillation amplitude (Fig.\ref{fig3}j) is overall akin to that of the rotating regime, the angular velocity (Fig.\ref{fig3}k) is approximately twice as large. This is because of  the greater active stress. The resulting steady state shapes of the core and shell  concurrently undergo rapid variations over time with near-average values (Fig.\ref{fig3}l), although the core one is slightly lower (meaning higher deformations) than that of the rotating regime (Fig.\ref{fig3}h) because of the larger flow in the shell.  

These results show that increasing activity leads to different motility regimes in which the spontaneous flow exhibits three well-defined structures: a double vortex in the translational regime, a single vortex in the rotating regime and a chaotic structure in the meandering  regime. Importantly, in each case the confined environment of the shell, combined with the perpendicular anchoring of the director  on both fluid interfaces, avoids the formation of topological defects altogether. 
This picture significantly changes if more than a single internal droplet is embedded in the outer active droplet. Indeed, as discussed in the introduction, a  pair of passive fluid droplets with strong anchoring dispersed in a larger outer active droplet must be, by necessity, topologically non-trivial.

{\bf Two-core active emulsion}. A two-core emulsion consists of an active nematic droplet encapsulating two passive isotropic droplets of equal size, initially placed side by side at the center of the system. Phase field parameters and initialisation are detailed in Methods and Supporting information.  The strong homeotropic surface anchoring nucleates a static disclination loop, with an odd hedgehog charge. In passive emulsions, the loop is static (Fig.\ref{fig1}b), which remains true for sufficiently weak activity.

\begin{figure*}[t!]
\includegraphics[width=2.0\columnwidth]{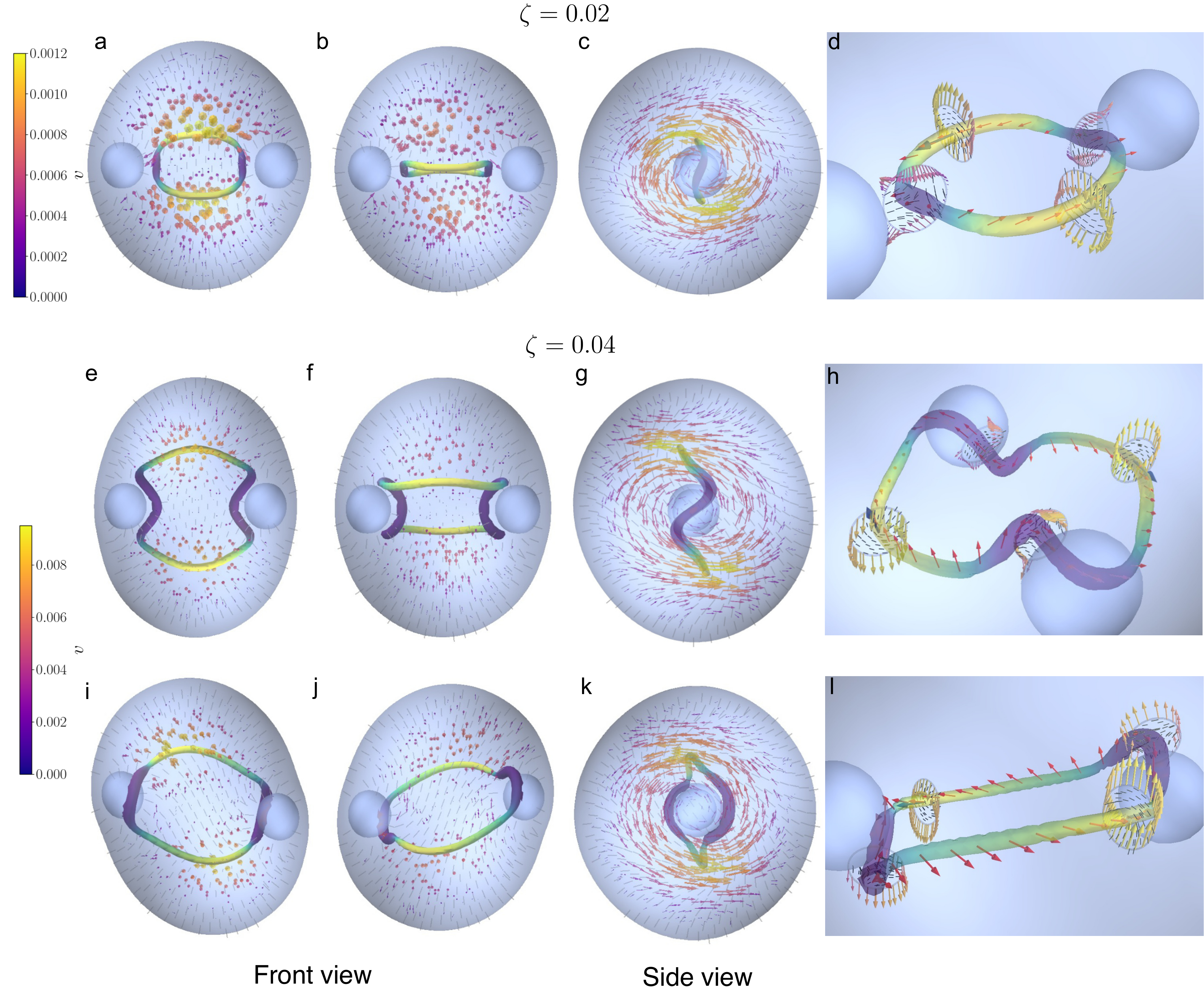}
\caption{\textbf{Extensile two-core emulsion with homeotropic anchoring: Rotating regime}. (a-b) Front view of typical structure of the director field (headless vectors) velocity field and disclination lines during time evolution of a two-core emulsion in the rotating regime (panels correspond to $t=4\times 10^{5}$ and $t=4.5\times 10^{5}$ respectively), for $\zeta=0.02$.  The loop disclination consists of two regions of $+1/2$ defect profiles (yellow segments), which smoothly convert into  regions
of two $-1/2$ profiles coasting the cores (purple segments)
by means of intermediate twist defects (green segments). (c) Velocity field in the longitudinal plane for the same configuration shown in (b). Panel (d) shows the director and the velocity field near the $+1/2$ and $-1/2$ defect profiles.
Panels (e-f) and (i-j) refer to the case $\zeta=0.04$ ($t=4\times 10^{5}$ and $t=4.5\times 10^{5}$, $t=5\times 10^{5}$ and $t=5.5\times 10^{5}$  respectively). Panels (g) and (k) show the side view of the velocity field for the same configuration in panels (e) and (i). Panels (h) and (l) show the director field (black headless vectors) and the velocity field around the disclination line for the same configurations. }
\label{fig4}
\end{figure*}

\begin{figure*}[ht!]
\includegraphics[width=2.0\columnwidth]{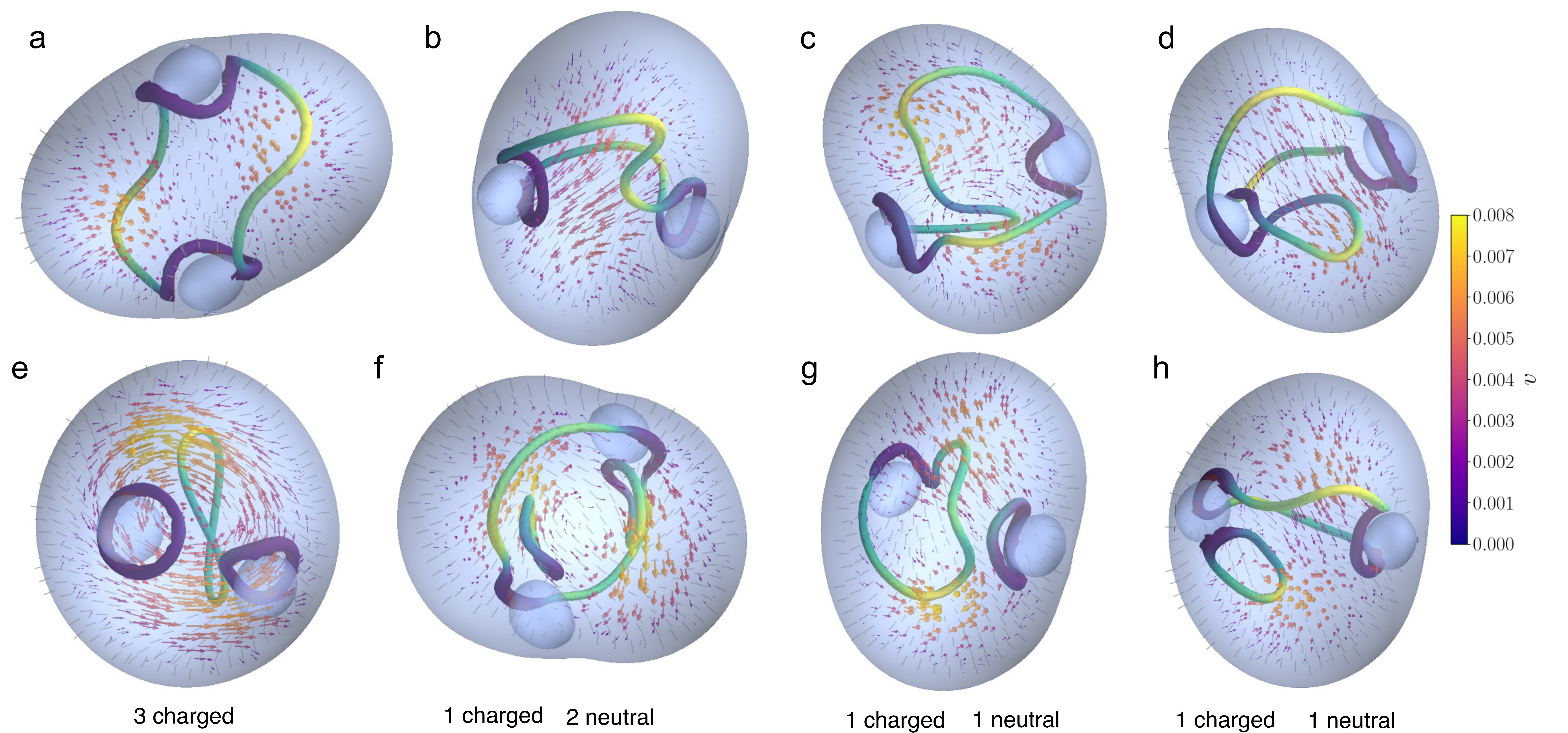}
\caption{\textbf{Extensile two-core emulsion with homeotropic anchoring: Chaotic regime.} (a-h) Different configurations of a two-core emulsion for $\zeta=0.08$. In this regime the charged loop departs from the spatio-temporally ordered flow states, continuously stretching and writhing into complex conformations (a-d). The continual stretching/shrinking and $+1/2$-driven threading dynamics eventually facilitates the loop to undergo splitting and recombination dynamics, which leads to multiple loops (e-h), with different possible combinations of charged and uncharged loops.}
\label{fig5}
\end{figure*}

We characterise the topology and local structure of these defects in terms of the behaviour of the unit vectors ${\bm{\Omega}}$ and ${\bf T}$, indicating  respectively the rotation vector --- which identifies the winding plane of the local director --- and the disclination line tangent (Fig.\ref{fig1}c) \cite{binysh2020,schimming2022,l_head}. More specifically,  monitoring the values of $\cos\beta=\bm{\Omega}\cdot{\bf T}$ allows for the identification of different local director profiles, such as twist-type ($\cos\beta=0$), $+1/2$ comet-shaped ($\cos\beta=1$) and $-1/2$ trefoil-shaped ($\cos\beta=-1$; note that $/pm 1/2$ refers to the topological charge of the in-plane defect profile).
Remarkably, such an inspection unveils a number of unique features in active two-core emulsions. At low activity ($\zeta=0.02$ or $Er \simeq 78$), the loop disclination consists of two segments of $+1/2$ defect profiles (yellow segments, Fig.\ref{fig4}a,b) that smoothly convert into  two $-1/2$ profile segments coasting the cores  (purple segments) by means of intermediate twist defects (green segments). The double parity transformation between $\mathbf{\Omega}$ and $\mathbf{T}$, in the terminology of ~\cite{copar2014}, is one of the possible characteristic patterns for an elliptical-shaped charged disclination loop (Supplementary Note 2, Fig.S1c). In contrast to neutral loops with approximately  constant $\mathbf{\Omega}$ along the contour (Supplementary Note 2, Fig.S1d), simple charged loops exhibit a $2\pi$ winding of $\mathbf{\Omega}$ along the loop, which enables the director to sample all orientations on the unit sphere (and carry a hedgeghog charge). By arbitrarily orienting the loop so that $\mathbf{T}$ turns anticlockwise, the observed +-+- profile corresponds to $\mathbf{\Omega}$ winding in the opposite sense. The charged -1/2 disclinations loops associated with nematic colloids ~\cite{weitz,copar2014} (Supplementary Note 2, Fig.S1a) have $\mathbf{\Omega}$ rotate in the same anticlockwise sense as $\mathbf{T}$, but with $\mathbf{T}$ and  $\mathbf{\Omega}$ oppositely oriented with respect to each other. All of these simple charged loops are topologically equivalent but differ by geometric transformations of the rotation axis of $\mathbf{\Omega}$, and relative orientation between $\mathbf{T}$ and $\mathbf{\Omega}$.
This pattern is also observed for a small defect loop in the passive case (Fig.\ref{fig1}b) as a consequence of the external environment attracting the colloids by the splayed confinement. However, extensile activity is known to favour twist defect profiles~\cite{shendruk2018,duclos2020} and self-propelling active backflows around +1/2 comet-shaped segments~\cite{binysh2020}. Therefore we interpret this +-+- non-trivial pattern as arising due to a competition between the preference of bulk active nematics for twist profiles and the anchoring, which pins the in-plane $-1/2$ profile close to the core and smoothly visits $+1/2$.

With respect to the active flow pattern, the rotation of the loop is powered by the active torque provided by the two azimuthally-oriented $+1/2$ self-propelled defect profiles (Fig.\ref{fig4}d). 
This feature is highlighted by the velocity field, which, in the vicinity of the loop, exhibits a unidirectional trend around $+1/2$  segments and a non-motile contractile-like dipolar flow  structure around $-1/2$ segments, globally producing a single clockwise vortex in the longitudinal plane (Fig.\ref{fig4}c and Supplementary Movie 4).

Increasing the activity ($\zeta=0.04$ or $Er\simeq 157$) strengthens the active back-flows, retaining the rotor-like loop dynamics but now with a stretched contour length (Supplementary Note 2, Fig.S2b) and `s'-shaped coiling of the -1/2 disclination segments around the passive cores (Fig.\ref{fig4}e-h). Here, the local writhe supports extended -1/2 segments to sweep out the director solid angle (Fig.\ref{fig4}h). In between the -1/2 profiles, the twist and +1/2 profiles have an approximately uniform $\mathbf{\Omega}$, suggesting that the contributions to the odd charge are spatially localised close to the inner droplets. The two azimuthal +1/2 segments power a coherent central vortex that, in turn, supports the loops rotation, all while axially pinned to the cores by the passive -1/2 segments.
With the +1/2 profiles located equidistant between the two inner cores, the flow state is achiral. However, this symmetry is unstable and eventually 
transitions into a chiral pattern (Fig.\ref{fig4}i-l). In this transition, the $+1/2$ profiles skew towards alternative inner emulsions (Fig.\ref{fig4}l), driving a central vortex that periodically switches the rotation axis (Fig.\ref{fig4}k and Supplementary Movie 5). Correspondingly, the -1/2 disclination segments transition from an `s' shape to a `c'-shaped conformation, which forms close to a full -1/2 circle from the side view. 


At yet larger activity ($\zeta=0.08$ or $Er\simeq 315$), the emergence of chaotic dynamics contributes rich dynamical and topological states (Fig.\ref{fig5} and Supplementary Movie 6) that resemble active living polymers~\cite{cates1987}. The charged loop departs from the spatio-temporally ordered flow states, instead continuously stretching and writhing into complex conformations (Fig.\ref{fig5}a-d). The departure from the simple elliptic loops into coiled and extended hairpin structures means that the non-trivial loop can depart from the two +1/2, two -1/2 segments. Indeed, examples can be seen with three +1/2 profiles (Fig.\ref{fig5}a-d), that locally `pinch' and drive the disclination loops into threaded dynamics. Loop `pinching' is associated with $+1/2$ segments due to their self-motile active backflows~\cite{binysh2020}. The continual stretching/shrinking and $+1/2$-driven threading dynamics eventually leads to splitting and recombination dynamics, generating multiple loops (Fig.\ref{fig5}e-h). Topological constraints restrict the number and nature of possible loop combinations. Since the two-core emulsions require only an odd bulk charge, configurations with 3-charged loops can be observed transiently (Supplementary Note 2, Fig.S2c). Charge neutral loops have no topological constraints on their number, and examples 
can be observed with one (Fig.\ref{fig5}g,h) or two (Fig.\ref{fig5}e,f) trivial loops accompanying the charged loop. In each case, loops are always observed to have -1/2 disclination profiles pinned to the passive cores, demonstrating that geometric constraints are crucial to the self-assembled defect organisation in these confined topological environments.

\begin{figure}[t!]
\includegraphics[width=0.991\columnwidth]{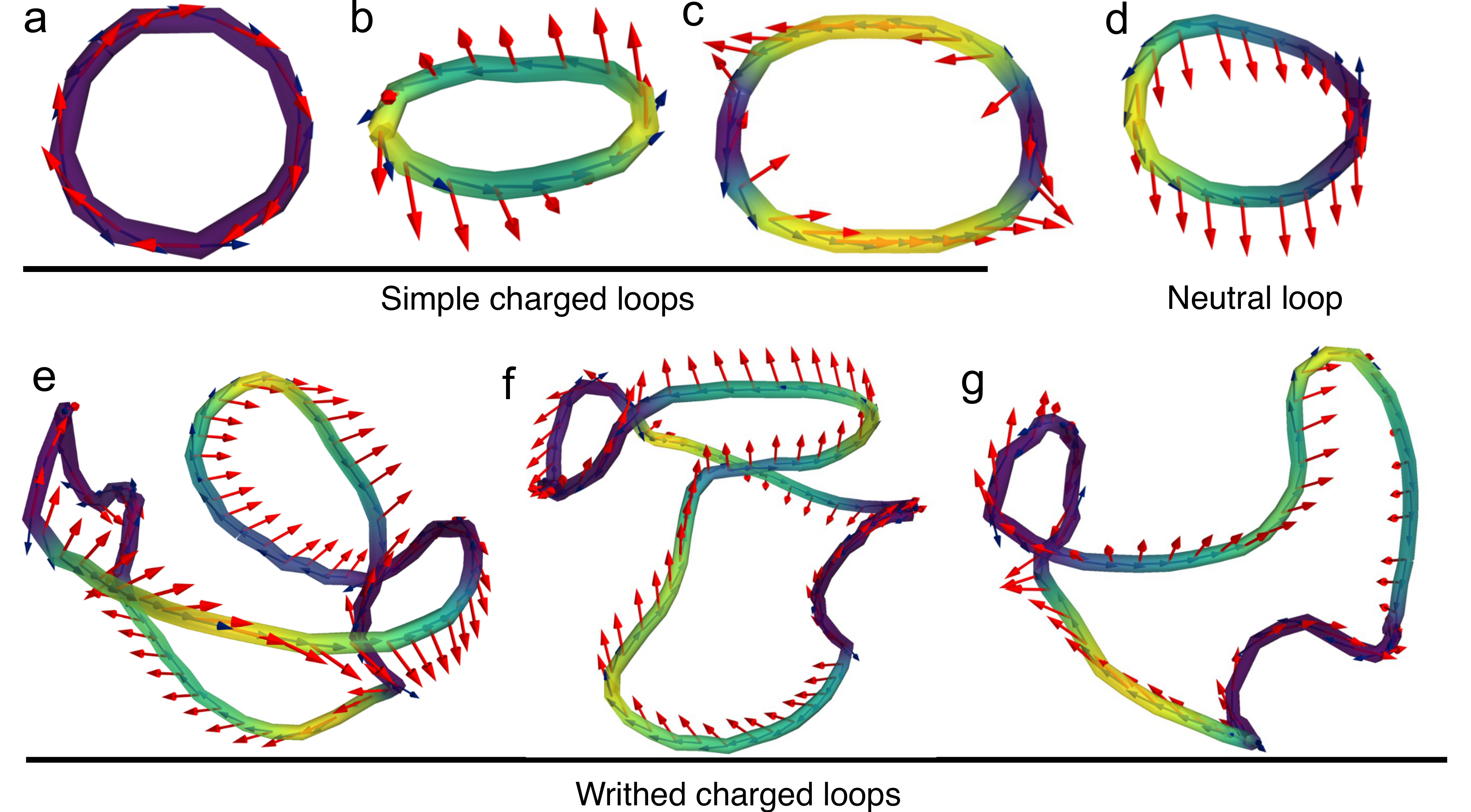}
\caption{\textbf{Types of disclination loops observed in two-core emulsions.} (a) Typical $-1/2$ disclination loop constrained to inner cores. (b) $+1/2$ -twist charged loop. (c) $+-+-$ charged loops observed for low activity. (d) Comet-twist neutral loop. (e-g) Writhed charged loops observed in the chaotic regime.  }
\label{fig6}
\end{figure}

In Fig.\ref{fig6} we summarise each type of disclination loops observed in these double inner emulsion states. For low activity only the +-+- loops are observed (Fig.\ref{fig6}c). This is favoured since each of the cores pins a non-motile $-1/2$ profiles. In the high activity regimes, topological transitions between combinations of odd numbers of charged loops and integer numbers of neutral loops are allowed. These  can additionally realise pure -1/2 disclination loops constrained to single cores (Fig.\ref{fig6}a, Fig.\ref{fig5}e,g) or bulk +1/2-twist charged loops (Fig.\ref{fig6}b, and Supplementary Note 2, Fig.S1b) when the emulsions are already occupied with -1/2 disclinations (Fig.\ref{fig6}a). More complex realisations of charged loops emerge when motile +1/2 profile segments are able to `pinch' loop segments. Despite varying curvature along the loop tangent, the rotation vector $\mathbf{\Omega}$ is observed to  slowly vary in the bulk fluid, and sharply turn near the passive cores, which for large loops is accompanied by local writhe (Fig.\ref{fig6}e-g). Due to the strong geometric confinement inside the inner emulsions, neutral loops are only seen as comet-twist loops (Fig.\ref{fig6}d), in contrast to the preference for twist-type loop seen in bulk active turbulence  (Supplementary Note 2, Fig.S1e) \cite{duclos2020}.



\section{Discussion}

In summary,  this work has studied the dynamical states of active nematic double emulsions, which provide an example of topological active matter that is experimentally realisable~\cite{MaaasPRL,review_autonomous_materials}. These emulsions consist of a large active nematic droplet containing a number of smaller passive droplets (cores) in its interior. The number of cores determines the global topology of the director field. 
The nematic textures associated to disclination loops contain features of line and point defects. Disclination loops are known to only display two types of topological point charge, even or odd. 
Even charge is the trivial element, corresponding to charge neutrality, whereas odd charge is the nontrivial element with a point charge texture~\cite{mermin1979}.
An emulsion with an \textcolor{black}{odd} number of passive cores is topologically trivial (so that a defect-free state is in principle possible), whereas one with an \textcolor{black}{even} number of cores is topologically non-trivial (so that there must be at least one disclination in the system).
By changing the number of cores and activity, it is therefore possible to study the physics emerging from the interplay between global topological constraint and dipolar active forces.

In the case of the topologically trivial single-core emulsion, low activity renders the system self-motile, whilst remaining defect-free. The steady state motility patterns are notable, because they encompass a transition between translational and rotational motion at higher activity. This transition was not previously observed in single 3D active nematic droplets:  there, tangential anchoring leads to only rotation, while helical trajectories  are only observed when including chiral contributions, either in the free energy~\cite{whitfield2017,carenza2019} or in the stress tensor~\cite{tjhung2017}. As further increasing activity yields chaotic dynamics, this double emulsion provides a remarkable example of a simple active matter system with a particularly rich dynamical behaviour.


When there are two internal passive cores, the double emulsion is now topologically nontrivial, because the normal anchoring at all surfaces induces the bulk liquid crystal pattern to have an odd topological charge. This two-core emulsion can spontaneously rotate, either regularly or in a chaotic fashion, according to the value of the activity. In the regularly rotating regime the steady state involves formation of a large, essentially planar, charged disclination loop. This loop has an interesting geometry, where two local $+1/2$ defect profiles  coexist with two other local $-1/2$ profiles. This +-+- pattern corresponds to a double revolution of the director field frame as the loop is traversed once, hence, as required by the boundary condition, the loop is topologically charged~\cite{copar2013,copar2014}. This charged loop is qualitatively different from those  in an active nematic droplet without passive cores~\cite{binysh2020}, which is also a topologically non-trivial system but where the local defect profile is more commonly twist. In our case, the $-1/2$ profile appears in the vicinity of the colloidal cores, in analogy with the profiles associated with Saturn rings around passive colloidal particles in nematics~\cite{copar2014}.

As the activity increases and the two-core emulsion enters the chaotic regime, the charged loop stretches and writhes in 3D; it also recombines topologically, for instance to create a charged loop around one of the cores and an uncharged loop in between the two cores. In this regime, the loop provides an intriguing example of a living active polymer~\cite{cates1987}, and it would be interesting in the future to pursue this analogy further.  These results suggest that the combination of active flows, plus inclusion-induced topological constraints and confinement-induced geometric constraints, can be utilised to design target disclination patterns and corresponding flow states. Geometric constraints allow control over the orientation of the rotation vector $\mathbf{\Omega}$, and the $\mathbf{\Omega}$ rotation plane in the case of simple elliptic charged loops. Systems with further numbers of inner emulsions may find higher order winding profiles of simple loops, and establish additional constraints to the chaotic conformational dynamics of writhed charged loops. 

The main message of this study is that double emulsions provide an example of self-assembled topological active material with tunable internal patterns. Topology is controlled at a global level via the number of passive cores included in the emulsion, whereas the patterns can be selected by tuning the value of the dipolar activity. The interplay between the emulsion topology and  activity leads to the emergence of various motility modes,  which are selectable in this framework. 
The system could  be self-assembled with double emulsions with microtubules and molecular motors, or other types of biomimetic or synthetic active matter, thereby providing an avenue to testing our predictions in experiments, paving the way for the development of novel, controllable functional materials. 

From the theoretical point of view, it would be of interest to extend the present study to incorporate pattern formation with more passive cores, and include chirality~\cite{whitfield2017,tjhung2017}, which is likely to lead to even richer topology and self-motile patterns~\cite{carenza2019}. Besides representing a novel type of topological active material and providing a rich playground for testing theories linking topology and pattern formation, active double emulsions  also have value biomimetically, as a toy model for a cell with a single centrosome or nucleus (a single-core emulsion), or a dividing cells with a mitotic spindle  pushing on two centrosomes (a two-core emulsion). Within this context, it would be of interest to study the minimal ingredients leading to ``cell motility''~\cite{tjhung}, ``nuclear rotation''~\cite{kumar2014} and ``cell division''~\cite{giomi2014}, which may provide a route for  future research aimed at self-assembling minimal cells~\cite{kohyama2022}. Here, we have considered extensile activity (e.g., due to microtubule and kinesin) and normal anchoring; we expect that addressing these fascinating questions on biomimetic systems will likely require supplementing our theory with actomyosin contractile, or mixed, activities and different types of boundary conditions at the droplet surfaces.

\section{Methods}
Here we outline the hydrodynamic model used in this work. We describe the physics of a liquid crystal double emulsion in terms of the following coarse-grained quantities: (i) the global fluid velocity ${\bf v}({\bf r}, t)$, (ii) a set of passive scalar phase field  $\phi_i({\bf r}, t)$ ($i=1,2$ for a single core emulsion and $i=1,2,3$ for the emulsion with two inner cores), capturing the density of each droplet and (iii) a tensor order parameter ${\bf Q}({\bf r},t)$ accounting for the ordering properties of a liquid crystal made of rod-like molecules. In the uniaxial approximation, $Q_{\alpha\beta}=q(n_{\alpha}n_{\beta}-\frac{1}{3}\delta_{\alpha\beta})$ (Greek subscripts denote Cartesian components), where ${\bf n}$ represents the local orientation of the liquid crystal molecules (often termed director)
and $q$ gauges the amount of local order, a quantity proportional to the
largest eigenvalue of ${\bf Q}$ ($0\leq q\leq 2/3$).

The ground state of the mixture is encoded in the following free energy density
\textcolor{black}{\begin{eqnarray}\label{free}
&f&=\sum_{i=1}^{N} \frac{a}{4}\phi_i^2(\phi_i-\phi_0)^2 + \frac{k}{2}\sum_{i=1}^{N}(\nabla\phi_i)^2  + \sum_{i,j,i<j}\epsilon\phi_i^2\phi_j^2\nonumber
\\&&+\frac{A_0}{2}\left(1-\frac{\chi(\bar\phi)}{3}\right)Q_{\alpha\beta}^2-\frac{A_0\chi(\bar\phi)}{3}Q_{\alpha\beta}Q_{\beta\gamma}Q_{\gamma\alpha}\nonumber\\&&+\frac{A_0\chi(\bar\phi)}{4}(Q^2_{\alpha\beta})^2+\frac{\kappa}{2}(\partial_{\gamma}Q_{\alpha\beta})^2\nonumber\\&&+W\sum_{i=1}^N\left[\partial_{\alpha}\phi_iQ_{\alpha\beta}\partial_{\beta}\phi_i\right]. 
\end{eqnarray}}
Here the first term, multiplied by the positive constant $a$, represents a double-well potential that ensures the existence of two coexisting minima at $\phi_i=\phi_0$ and $\phi_i=0$. For an emulsion with a single core $N=2$, while $N=3$ for an emulsion with two passive cores. 
In the first case $\phi_1=\phi_0$ inside the inner core and $\phi_1=0$ outside the core. The outer droplet is represented by an additional phase field $\phi_2$, with $\phi_2=0$ inside the droplet and $\phi_2=\phi_0$ outside.
For an emulsion with two passive cores, an additional phase field with the same properties as $\phi_1$ is used.
The second term in Eq.(\ref{free}), multiplied by the elastic constant $k$, controls the interfacial energy. 
Both constants $a$ and $k$ fix
the surface tension $\sigma=\sqrt{{8ak/9}}$ and the interface thickness $\xi_{\phi}=\sqrt{2k/a}$ of the droplets.
The third contribution is a soft-core repulsion, whose magnitude is
controlled by the positive constant $\epsilon$. This term, necessary to prevent droplet merging, mimics the effect produced, at a mesoscale level, by a surfactant adsorbed onto the droplet interfaces. The bulk properties of the liquid crystal are described by a fourth order expansion of the Q-tensor
(summation over repeated indices is assumed), where $A_0$ is a positive constant and $\chi(\bar\phi)$, with $\bar\phi=\sum_{i=1}^N\phi_i$, controls the isotropic-liquid crystal transition, which occurs when $\chi(\bar\phi)>\chi_{cr}=2.7$ \cite{degennes}. Following previous works \cite{carenza2019,sulaiman}, we set $\chi=\chi_0+\chi_s\bar\phi$, where  $\chi_0$ and $\chi_s$ control the boundary of the coexistence
region. Note that $\chi$ depends exclusively on $\bar\phi$, since the liquid crystal is confined solely within the layer where $\bar\phi=0$ (see Fig.\ref{fig1}). The energetic cost due to liquid crystal distortions is gauged by the gradients of $Q_{\alpha\beta}$ (where $\kappa$ is the elastic constant) while the anchoring 
of the director at the droplet interface is described by the last term where $W$ is the anchoring strength. If $W>0$ the anchoring is tangential, otherwise it is perpendicular. Here we only focus on the latter.

The dynamics of the scalar fields $\phi_i$  obeys a Cahn-Hilliard equation
\begin{equation}\label{cahn_eqn}
\partial_t \phi_i+{\bf v}\cdot{\nabla\phi_i}=M\nabla^2\mu_i,
\end{equation}
where $M$ is the mobility and $\mu_i=\frac{\delta{\cal F}}{\delta\phi_i}$
is the chemical potential, with ${\cal F}=\int_V f d{\bf r}$.

The time evolution of the ${\bf Q}$ tensor is governed by the Beris-Edwards equation \cite{beris}
\begin{equation}\label{beris_eqn}
(\partial_t + {\bf v}\cdot\nabla){\bf Q}-{\bf S}({\bf W},{\bf Q})=\Gamma{\bf H},
\end{equation}
where $\Gamma$ is a collective rotational diffusion constant and ${\bf H}$ is the molecular field, which is given by
\begin{equation}
{\bf H}=-\frac{\delta {\cal F}}{\delta {\bf Q}}+({\bf I}/3)Tr\frac{\delta {\cal F}}{\delta {\bf Q}},
\end{equation}
with ${\bf I}$ unit matrix. 
The first term on the left-hand side of Eq.\ref{beris_eqn} represents the material derivative accounting for the time dependence of a quantity advected by the fluid velocity ${\bf v}$. The second term is further contribution accounting for rotation and  stretching  of the rod-like molecules of the liquid crystals due to flow gradients \cite{beris}, and is given by
\begin{eqnarray}
S({\bf W},{\bf Q})&=&(\xi{\bf D}+{\bm\omega})({\bf Q}+{\bf I}/3)+
(\xi{\bf D}-{\bm\omega})({\bf Q}+{\bf I}/3)\nonumber\\&&
-2\xi({\bf Q}+{\bf I}/3)Tr({\bf Q}{\bf W}).
\end{eqnarray}
Here $Tr$ denotes the tensorial trace, while ${\bf D}=({\bf W}+{\bf W}^T)/2$ and ${\bm\omega}=({\bf W}-{\bf W}^T)/2$ are the symmetric and anti-symmetric part of the velocity gradient tensor $W_{\alpha\beta}=\partial_{\beta}v_{\alpha}$. The role of the constant $\xi$ is twofold. On the one hand, 
it determines the aspect ratio of the liquid crystal molecules; it is  positive for rod-shaped molecules and negative for disk-like ones. On the other hand, it 
controls  whether the director field is flow aligning or flow tumbling under shear. In such conditions, at the steady state the director would align with the flow gradient at
an angle $\theta$ such that $\xi cos(2\theta) = (3q)/(2 + q)$ \cite{degennes}, whose real solutions (corresponding to a flow aligning regime) are obtained if $\xi\geq 0.6$.

Finally, the fluid velocity ${\bf v}$ is ruled by the incompressible Navier-Stokes equations
\begin{equation}\label{cont_eqn}
\nabla\cdot{\bf v}=0,
\end{equation}
\begin{equation}\label{nav_stok_eqn}
\rho(\partial_t{\bf v}+{\bf v}\cdot \nabla{\bf v})=-\nabla p + \nabla\cdot ({\bm \sigma}^{visc}+{\bm \sigma}^{lc}+{\bm \sigma}^{int}+{\bm \sigma}^{act}),
\end{equation}
where $\rho$ is the fluid density and $p$ is the hydrodynamic pressure. The viscous contribution is given by
\begin{equation}
\sigma_{\alpha\beta}^{visc}=\eta(\partial_{\alpha}v_{\beta}+\partial_{\beta}v_{\alpha}),
\end{equation}
where $\eta$ is the shear viscosity of the fluid, while the elastic one due to liquid crystal deformations reads
\begin{eqnarray}
\sigma_{\alpha\beta}^{lc}=&&-\xi H_{\alpha\gamma}(Q_{\gamma\beta}+\frac{1}{3}\delta_{\gamma\beta})-\xi(Q_{\alpha\gamma}+\frac{1}{3}\delta_{\alpha\gamma})H_{\gamma\beta}\nonumber\\
&&+2\xi(Q_{\alpha\beta}-\frac{1}{3}\delta_{\alpha\beta})Q_{\gamma\mu}H_{\gamma\mu}+Q_{\alpha\gamma}H_{\gamma\beta}-H_{\alpha\gamma}Q_{\gamma\beta}\nonumber\\
&&-\partial_{\alpha}Q_{\gamma\mu}\frac{\partial f}{\partial(\partial_{\beta}Q_{\gamma\mu})}.
\end{eqnarray}
A further contribution stems from interfacial stress between the active and the passive phase, and reads as follows,
\begin{equation}
\sigma^{int}_{\alpha\beta}=\sum_{i=1}^N\left[\left(f-\phi_i\frac{\delta{\cal F}}{\delta\phi_i}\right)\delta_{\alpha\beta}-\frac{\partial f}{\partial(\partial_{\beta}\phi_i)}\partial_{\alpha}\phi_i\right].
\end{equation}
The last term is the active stress given by \cite{marchetti,hatwalne}
\begin{equation}
\sigma_{\alpha\beta}^{act}=-\zeta Q_{\alpha\beta},
\end{equation}
where $\zeta$ is the active parameter, positive for extensile materials and negative for contractile ones.

As in previous works \cite{tjhung2015,carenza2022,carenzasoftchannel,Negro_2019method}, we use a 3D hybrid lattice Boltzmann method (see SI\cite{SI}), which solves Eq.\ref{cahn_eqn} and Eq.\ref{beris_eqn} via a finite difference scheme while Eqs.\ref{cont_eqn} and Eq.\ref{nav_stok_eqn} by a standard LB approach. A complete list of parameters values is provided in the Supplementary Note 1. 

The angular velocity of the droplets has been computed as:
$ \omega = \int \text{d}\mathbf{r} \phi \dfrac{\Delta \mathbf{r} \times \Delta \mathbf{v}}{|\Delta \mathbf{r}|^2},$ where $\Delta \mathbf{r}= \mathbf{r} - \mathbf{R}$ and $\Delta \mathbf{v}= \mathbf{v} - \mathbf{V}$, being $\mathbf{R}$ and $\mathbf{V}$ respectively the position and the velocity of the center of mass of the droplets. 

The droplets deformations have been computed as $d_{max}/d_{min}$,  with $d_{max} \geqslant d_{min}$ the eigenvalues of the positive-definite Poinsot matrix associated to the droplet \cite{carenza2019,carenzaphysicaAchol}.

\subsection{Defect analysis}

Characterising disclination loops utilises the disclination density tensor, proposed by Schimmings and Vi\~nals \cite{schimming2022}. The tensor is constructed from derivatives of the Q-tensor
\begin{align}    D_{ij}=\epsilon_{i\mu\nu}\epsilon_{jlk}\partial_l Q_{\mu\alpha}\partial_k Q_{\nu\alpha}
\end{align}
where $i,j,k,\alpha,\mu,\nu$ are tensor indices with applied summation convention. The usefulness of this form comes from the interpretation as the dyad composing of the local line tangent $\mathbf{T}$ and the rotation vector $\mathbf{\Omega}$
\begin{align}
    \label{eq:Dtensordecomposition}
    D_{ij} = s(\mathbf{r})\mathbf{\Omega}_{i}\mathbf{T}_j,
\end{align}
where $s(\mathbf{r})$ is a positive scalar field that is maximum at the disclination core. In this work, we identify defects as isosurfaces where $s(\mathbf{r})=0.033$. To extract $\mathbf{\Omega}$ and $\mathbf{T}$,  we use the methods outlined in \cite{schimming2022}, ensuring that the vectors are continuous along the loop and have the correct relative sign (set by sgn$(\mathbf{\Omega}\cdot\mathbf{T})=$sgn(Tr($D_{ij}$))).
The contour lengths are found by grouping the disclination cells $s(\mathbf{r})\geq0.033$ into an ordered sequence of points that form a polymer loop, followed by summation of bond lengths.   

All three-dimensional visualisations use the Mayavi library \cite{ramachandran2011mayavi}.


\section*{Competing interests}
The authors declare no competing interests.

\section*{Acknowledgements}
Part of the numerical simulations were performed on the Dutch national e-infrastructure with the support of SURF through Grant 2021.028 for computational time (L. N. C. and G. N.), on ReCas HPC-Cluster in Bari (Italy) (G.N.), and on Galileo100 at CINECA, Italy (G.N.).  We acknowledge funding from MIUR Project No. PRIN 2020/PFCXPE. A. T. acknowledges CNR for funding his visit at the University of Edinburgh within the Short Term Mobility Program 2023.
This research has received funding from the European Research Council (ERC) under the European Union’s Horizon 2020 research and innovation programme (Grant agreement No. 851196). 
For the purpose of open access, the author has applied a Creative Commons Attribution (CC BY) licence to any Author Accepted Manuscript version arising from this submission. 

\bibliographystyle{unsrt}
\bibliography{bibliography.bib}


\end{document}